\documentclass[prb,twocolumn,floatfix,nobibnotes,footinbib,superscriptaddress,
showpacs]{revtex4}
\usepackage[latin1]{inputenc}
\usepackage{amsmath}
\usepackage{graphicx}
\usepackage{psfrag}
\begin{document}
\title{Dual fermion approach to the two-dimensional Hubbard model:
Antiferromagnetic fluctuations and Fermi arcs}

\author{A. N. Rubtsov}
\affiliation{Department of Physics, Moscow State University,
119992 Moscow, Russia }
\author{M. I. Katsnelson}
\affiliation{Institute for Molecules and Materials, Radboud
University, 6525AJ Nijmegen, The Netherlands}
\author{A. I. Lichtenstein}
\affiliation{Institute of Theoretical Physics, University of
Hamburg, 20355 Hamburg, Germany}
\author{A. Georges}
\affiliation{Centre de Physique Theorique, CNRS, Ecole
Polytechnique, 91128 Palaiseau Cedex, France}


\begin{abstract}
We present an efficient diagrammatic method to describe nonlocal
correlation effects in lattice fermion Hubbard-like models, which
is based on a change of variables in the Grassmann path integrals.
The new fermions are dual to the original ones and correspond to
weakly interacting quasiparticles in the case of strong local
correlations in the Hubbard model. The method starts with
dynamical mean-field theory as a zeroth-order approximation and
includes non-local effects in a perturbative way. In contrast to
cluster approaches, this method utilizes an exact transition to a
dual set of variables. It therefore becomes possible to treat
vertices of an effective {\it single-impurity} problem
as small parameters. This provides a very efficient interpolation
between band-like weak-coupling and atomic limits. The method is
illustrated on the two-dimensional Hubbard model. The
antiferromagnetic pseudogap, Fermi-arc formations, and
non-Fermi-liquid effects due to the van Hove singularity are
correctly reproduced by the lowest-order diagrams. Extremum
properties of the dual fermion approach are discussed in terms of
the Feynman variational principle.
\end{abstract}


\maketitle

\section{Introduction}

One of the most successful theories of strongly correlated
fermions on a lattice is dynamical mean-field theory
(DMFT)\cite{dmft}. Physically, this approach treats the local spin
and orbital fluctuations of the correlated electrons in a correct
self-consistent way , while the spatial intersite correlations on
the lattice are neglected. The non-perturbative DMFT approach is
successful, because a number of the most important correlation
effects are indeed related to local fluctuations. For example,
DMFT describes correctly such phenomena, as the local moment
formation in itinerant magnets \cite{Moriya}, some aspects of
Kondo physics \cite{hewson}, and the Mott insulator-to-metal
transition on a lattice with a large connectivity in
high-dimensional materials \cite{dmft}.

On the other hand, there is increasing evidence that the
non-locality of spatial correlations plays an important role,
particularly for the Luttinger liquid physics of low-dimensional
correlated systems\cite{Giamarchi}, d-wave pairing in quasi
two-dimensional cuprates\cite{anderson2, dwave}, and non
Fermi-liquid behavior due to van-Hove singularities in
two-dimensional systems \cite{VH,fRG,vanHove}. Moreover,
angle-resolved photoemission spectra of three-dimensional
ferromagnetic iron shows appreciable k-dependent self-energy
effects \cite{Claessen}.

The most obvious generalizations of DMFT that takes into account
the short-range non-local fluctuations are the so-called cluster
DMFT approximations, in real or k-space \cite{cdmft,dca}. In these
methods, correlations are assumed to be localized within a cluster
including several lattice sites. Cluster methods do catch the
basic physics of $d$-wave pairing and anti-ferromagnetism in
high-T$_c$ superconductors\cite{cdmftdw,plaquette} and the effects
of inter-site Coulomb interaction in transition-metal oxides
\cite{ti2o3}. At the same time, the complicated k-dependence of
the self-energy close to the Fermi surface, giving rise to
Luttinger liquid formation is related to long-range fluctuations
and therefore cannot be described within cluster approaches. For
the same reason, cluster methods hardly can handle the effects due
to van-Hove singularities or nesting \cite{VH,vanHove}. Another
drawback of the cluster methods is that the specific choice of the
cluster and corresponding self-consistency condition is not
unique. Different self-consistency conditions (e.g. DCA\cite{dca}
and free-cluster CDMFT\cite{cdmft}) or periodization schemes (e.g.
self-energy and cumulant periodization \cite{plaquette}) can
result in physically different solutions. For example, the
critical temperature of the $d$-wave superconducting transition of
the doped Hubbard model is different in DCA calculations
\cite{dca} and for then $2\times 2$ free cluster \cite{plaquette}.

The present paper is devoted to an alternative extension of DMFT,
which operates with a {\it single-site} impurity problem and
treats spatial nonlocality in a diagrammatic way.

Let us first recall the key DMFT equations. Formally, the assumption
of local correlations means that the environment of a correlated
atom can be replaced with a Gaussian effective medium. Consequently,
the lattice problem reduces to the impurity problem. The later is
described by the effective impurity action
\begin{equation}\label{Impurity0}
S_{imp}=S_{at}+\sum_{\omega,\sigma} \Delta_ \omega
c^*_{\omega,\sigma} c_{\omega,\sigma},
\end{equation}
where $S_{at}$ is an action of the isolated or bare atom, and the
second term is the hybridization due to the rest of the lattice.
An important property of the DMFT approach is that this
hybridization function has non-trivial frequency dependence, so
that the approximation catches the physics of local fluctuations
of spin, charge, and orbital degrees of freedom. For example,it is
vital, for the description of Kondo physics \cite{hewson}.

It is obvious that the impurity problem is much simpler than the
original lattice one. Nowadays, a number of numerically efficient
impurity solvers are available. In particular, these solvers allow
one to calculate the Green's function of the impurity problem
$g_{\omega, \sigma}$ on the Matsubara frequencies axis. This is
the only property of the impurity problem entering in the DMFT
self-consistent equations. The DMFT approximation for the Green's
function of the initial lattice problem corresponds to the
following expression
\begin{equation}\label{G0}
    G_{\omega k \sigma}^{DMFT}=
    \frac{1}{g_{\omega, \sigma}^{-1}+\Delta_{\omega,
    \sigma}-\epsilon_k}.
\end{equation}
One can see from this equation that the self-energy is local in
DMFT, since the momentum dependence of $\epsilon_k$ is not
renormalized. The hybridization function $\Delta$ satisfies the
self-consistentency condition of DMFT,
\begin{equation}\label{cond0}
    G^{DMFT}_{r=0, \omega, \sigma}=g_{\omega, \sigma},
\end{equation}
where $G_{r=0}=N^{-1} \sum_k G_{k}$ is the local part of the
Green's function (\ref{G0}) of the lattice with $N$ sites.

In order to understand the main idea of the present work, let us
first describe in a simple way the DMFT condition (\ref{cond0}).
If we consider the case of a truly Gaussian system then the DMFT
approach becomes exact. For this case, equation (\ref{cond0}) is
trivial. Indeed, to obtain the impurity problem for the site $j$,
one integrates out truly Gaussian degrees of freedom for other
sites. This exact procedure does not change the properties of the
electron motion at the site $j$, so the local part of the Green's
function before integration must equal the Green's function after
the integration, $G_{R=0}=g$. Turning back to the general case of
a non-Gaussian ensemble, we note that among different properties
of the impurity model, the DMFT scheme uses only the local
Greens's function $g_{\omega\sigma}$. Once $g_{\omega\sigma}$ is
known, the approximation does not differ between Gaussian and
non-Gaussian cases. Therefore, if a certain equation for
$g_{\omega\sigma}$ is established for the Gaussian limit, it must
also remain valid for the general case.

As it follows from the previous discussion, the DMFT equations are
essentially the formulae for the Gaussian limit, renormalized in
terms of the Green's function of the impurity problem. It turns
out that the resulting theory works well, not only in the case of
weakly interacting systems, but also in the atomic limit case,
which is very different from a Gaussian system. A good
interpolation between the two different limits is a key advantage
of the DMFT approach.

Starting with the above interpretation of DMFT, it is natural to
discuss a possible extension of this theory. Such an extension
should be based on the perturbation series near the Gaussian
limit, renormalized in terms of the impurity problem. The
lowest-order term of such a theory should restore the DMFT result,
whereas higher-order corrections would describe spatial
non-locality. A properly constructed theory of this kind would
describe both short- and long-range fluctuations and will not
suffer from the periodization problems of cluster DMFT.

Unfortunately, the straightforward construction of such an
extension meets serious difficulties. The problem is that the
extension is not unique. Beyond DMFT, there are many ways to
choose the renormalization procedure, to define the hybridization
function for the impurity problem and other quantities. One can
formulate the major requirements for the desirable non-local
correlated theory, they include:
\begin{itemize}
    \item at least in the Gaussian and atomic limits, the theory
    should become a regular series around DMFT, with an explicit
    small parameter;
    \item the basic conservation laws should be fulfilled in the
    theory;
    \item the choice of hybridization function should be optimal,
   in a certain sense;
    \item there should be good practical convergence of the
    series: the leading corrections should capture most of the
    non-local physics;
    \item last but not least, the equations of the theory must be
    easy enough for practical calculations.
\end{itemize}

There have been several previous attempts to construct a proper
theory of this kind  \cite{Tos07,Kus06,Sle06}. These approaches
require a solution of ladder-like integral equations for the
complete vertex $\Gamma$ and the subsequent use of the
Bethe-Salpeter equation to obtain the Green's functions. The first
step exploits the vertex part of the effective impurity
problem, whereas the second step uses just the bare interaction
parameter $U$. We do not know of detailed tests of these
approaches \cite{Tos07,Kus06,Sle06}, but we suspect that the
presence of bare $U$ in the theory makes it suitable for the
metallic phases only. We also note that ladder-like integral
equations are hard for practical calculations.

In this paper, we describe in detail a formalism fulfilling all
the criteria from the above list. A preliminary version of this
method was published in Ref. \cite{Brief}. The method is based on
the transition to the new set of variables, called the dual
ensemble. The procedure utilizes a Hubbard-Stratonovich
transformation for the Gaussian part of the action. Several years
ago, this trick was first proposed for classical fluctuation
fields \cite{classic}. For a strong coupling expansion of the
Hubbard model around the atomic limit without hybridisation
function, the equivalent Hubbard-Stratonovich transformation has
been proposed in different papers \cite{Sarker,Tremblay}. A
similar procedure for fermions with general non-local interactions
have been discussed recently \cite{Kot04}. Also we would like to
mention a much earlier work \cite{GeorgesClass} for classical
fields. Although, it used a different formalism
\cite{GeorgesClass}, the resulting diagram series resembles ours.

The paper is organized as follows. Section II is devoted to the
general theoretical framework. Section III describes the
application of the non-local theory to the problem of the
antiferromagnetic pseudogap and the formation of Fermi arcs in the
two-dimensional Hubbard model for high-temperature superconducting
cuprates. In the Appendix A we discuss 
how the many-particle excitations for the initial and dual system are related.
 In the Appendix B the functional minimization derivation
of the self-consistent DMFT condition is discussed.

\section{Dual fermion formalism: beyond DMFT}

\subsection{Definitions}

We start from the two dimensional Hubbard model with the
corresponding imaginary-time action
\begin{equation}\label{Hubbard}
    S[c,c^*]=\sum_{\omega k \sigma}
    \left(\epsilon_k-\mu-i \omega\right) c^*_{\omega k \sigma} c_{\omega k \sigma} + U \sum_i \int_0^\beta n_{i\uparrow\tau} n_{i\downarrow\tau} d\tau.
\end{equation}
Here $\beta$ and $\mu$ are the inverse temperature and chemical
potential, respectively, $\omega=(2 j+1) \pi/\beta, j=0,\pm 1,
...$ are the Matsubara frequencies, $\tau$ is imaginary time,
$\sigma=\uparrow, \downarrow$ is the spin projection. The bare
dispersion law is $\epsilon_k=-2 t (\cos k_x+\cos k_y)$, $c^*, c$
are Grassmanni variables, $n_{i \sigma \tau}=c^*_{i \sigma \tau}
c_{i \sigma \tau}$, where the indices $i$ and $k$ label sites and
quasi-momenta.

In the spirit of the DMFT, we introduce a single-site reference
system (an effective impurity model) with the action
\begin{equation}\label{Impurity}
S_{imp}=\sum_{\omega,\sigma} (\Delta_ \omega-\mu-i \omega)
c^*_{\omega,\sigma} c_{\omega,\sigma} + U\int_0^\beta n_{\uparrow
\tau} n_{\downarrow \tau}  d\tau ,
\end{equation}
where $\Delta_\omega$ is a yet undefined hybridization function
describing the interaction of the effective impurity with a bath.
We assume that all properties of the impurity problem such as
single-particle Green's function $g_w$, and higher momenta can be calculated.
In particular, we will use the forth-order vertex 
$\gamma^{(4)}_{1234}=g^{-1}_{11'} g^{-1}_{22'} (\chi_{1'2'3'4'}-g_{1'4'} g_{2'3'}+g_{1'3'} g_{2'4'}) g^{-1}_{3'3} g^{-1}_{4'4}$ (here, $\chi$ is a two-particle Green's
function of the impurity problem, and indices stand for a combination of $\sigma$ and $\omega$, for example $g_{11'}$ means $g_{\sigma_1, \omega_1, \sigma_{1'}, \omega_{1'}}$).

 Our goal
is to express the Green's function $G_{\omega k}$ and other properties
of the lattice problem of Eq.(\ref{Hubbard}) via the 
quantities for the impurity problem.

\subsection{Dual variables: exact formulas} \label{ExactFormulas}

Since $\Delta$ is independent of $k$, the lattice action
(\ref{Hubbard}) can be represented in the following form
\begin{equation}\label{Hubbard2}
    S[c,c^*]=\sum_i S_{imp}[c_i, c_i^*]-\sum_{\omega  k \sigma}
    (\Delta_\omega-\epsilon_k) c^*_{\omega k \sigma} c_{\omega k
    \sigma}.
\end{equation}

\begin{widetext}
We utilize a dual transformation to a set of new Grassmann
variables $f, f^*$. The following identity
\begin{equation}\label{Gauss}
    e^{ A^2 c^*_{\omega k \sigma} c_{\omega k \sigma} }=
    \left(\frac{A}{\alpha}\right)^2
\int e^{- \alpha (c^*_{\omega k \sigma} f_{\omega k \sigma} +
f^*_{\omega k \sigma} c_{\omega k \sigma}) - \alpha^2 A^{-2}
f^*_{\omega k \sigma} f_{\omega k \sigma}}  d f^*_{\omega k
\sigma} d f_{\omega k \sigma},
\end{equation}
is valid for arbitrary complex numbers $A$ and $\alpha$. We chose
$A^2=(\Delta_\omega-\epsilon_k)$ for each set of indices $\omega,
k, \sigma$. The quantity $\alpha$ remains yet unspecified, but we
require it to be dispersionless; $\alpha=\alpha_{\omega, \sigma}$.

With this identity, the partition function of the lattice problem
$Z=\int  e^{-S[c,c^*]} {\cal D} c^* {\cal D} c$ can be presented
in the form $Z=\int \int e^{-S[c,c^*,f,f^*]} {\cal D} f^* {\cal D}
f {\cal D} c^* {\cal D} c$, where

\begin{equation}\label{Scf}
\begin{array}{c}
    S[c,c^*,f,f^*]=-\sum_{\omega k} \ln \left(\alpha_{\omega \sigma}^{2}
(\Delta_\omega-\epsilon_k)\right) + \sum_i S_{imp}[c_i, c_i^*]+\\
\\ +\sum_{\omega k \sigma}
     \left[\alpha_{\omega \sigma} (f^*_{\omega k\sigma} c_{\omega k\sigma} + c^*_{\omega k \sigma} f_{\omega k \sigma})
    +\alpha_{\omega \sigma}^2 (\Delta_\omega-\epsilon_k)^{-1}
    f^*_{\omega k \sigma} f_{\omega k \sigma}\right].
\end{array}
\end{equation}

As a  next step, we establish an exact relation between the
Green's function of the initial system $G_{\tau-\tau', i-i'}=-< T
c_{\tau i} c^*_{\tau' i'} >$ and that of the dual system
$G^{dual}_{\tau-\tau', i-i'}=- <T f_{\tau i} f^*_{\tau' i'}>$. To
this aim, we can replace $\epsilon_k \to \epsilon_k+\delta
\epsilon _{\omega k}$ with a differentiation of the partition
function with respect to $\delta \epsilon _{\omega k}$. Since we
have two expressions for the action (\ref{Hubbard}) and
(\ref{Scf}), one obtains
\begin{equation}\label{exact}
    G_{\omega, k}= (\Delta_\omega-\epsilon_k)^{-1} \alpha_{\omega \sigma}
    G^{dual}_{\omega, k} \alpha_{\omega \sigma} (\Delta_\omega-\epsilon_k)^{-1} +(\Delta_\omega-\epsilon_k)^{-1}.
\end{equation}
Similar relations hold also for higher-order momenta, as Appendix A describes.


The crucial point is that the integration over the initial
variables $c^*_i, c_i$ can be performed separately for each
lattice site, since $\alpha$ is local and $\sum_k \left( f^*_k c_k
+ c^*_k f_k \right) = \sum_i \left( f^*_i c_i + c^*_i f_i
\right)$. For a given site $i$, one should integrate out $c^*_i,
c_i$ from the action that equals
\begin{equation}\label{Ssite}
    S_{site}[c_i,c_i^*,f_i,f_i^*]=S_{imp}[c_i,c^*_i]+\sum_\omega
\alpha_{\omega \sigma} (f^*_\omega c_\omega + c^*_\omega
f_\omega).
\end{equation}
We finally obtain an action $S$ depending on the new variables $f,
f^*$ only;
\begin{equation}\label{Sf}
    S[f,f^*]=-\sum_{\omega k} \ln \left(\alpha_{\omega \sigma}^{-2}
(\Delta_\omega-\epsilon_k)\right)-\sum_i \ln z^{imp}_i+
\sum_{\omega k \sigma} \alpha_{\omega \sigma} \left((\Delta_\omega
- \epsilon_k)^{-1}
    +g_\omega\right) \alpha_{\omega \sigma} f^*_{\omega k \sigma} f_{\omega
    k \sigma} + \sum_i V_i,
\end{equation}
where $z^{imp}_i=\int e^{-S_{imp}[c^*_i,c_i]} {\cal D} c_i^* {\cal
D} c_i$, and the dual potential $V_i\equiv V[f^*_i,f_i]$ is defined
from the expression
\begin{equation}\label{defineV}
\int e^{-S_{site}[c^*_i,c_i, f^*_i, f_i]} {\cal D} c_i^* {\cal D}
c_i=z^{imp}_i e^{\sum_{\omega \sigma} \alpha_{\omega \sigma}^2
g_{\omega} f^*_{\omega i \sigma} f_{\omega i \sigma} - V[f_i,
f^*_i]}.
\end{equation}
\end{widetext}
The Taylor series for $V[f_i, f^*_i]$ can be obtained from the
expansion of this definition in powers of $f_i, f^*_i$. One can see
that (\ref{defineV}) defines $V$ in such a way that this series
starts from the quartic term, $\propto f^*f^* f f$. Later on we
take, for convenience:
\begin{equation}\label{definealpha}
    \alpha_{\omega \sigma}=g_\omega^{-1},
\end{equation}
as it gives a particularly simple form of $V$. In this case the
leading term in $V$ is $-\frac{1}{4}  \gamma^{(4)}_{1234} f^*_1 f^*_2 f_3 f_4$. Further Taylor
series terms yield similar combinations including $\gamma^{(n)}$
of higher orders.

Thus we see that in the dual action, the interaction terms remain
localized in space, but are they non-local in imaginary time,
since, for example, $\gamma^{(4)}$ depends on the three
independent Matsubara frequencies. Except for this point, the
action (\ref{Sf}) formally resembles (\ref{Hubbard}).

There is a point which is worthwhile to discuss here: one can
formally apply the transformation (\ref{Gauss}) with some new
hybridization function to the dual system (\ref{Sf}), and thus
obtain a sequence of changes to new variables. It is useless,
however, since mathematically, these transformations form a group.
It is easy to show that any sequence of the transformations
(\ref{Gauss}) corresponds to a single change of variables with a
certain $\Delta$. Moreover, there is an {\it inverse} change of
variables, that allows to obtain $S[c, c^*]$ back from the $S[f,
f^*]$. It is given just by Eq.(\ref{Gauss}) with $A$ replaced with
$\alpha A^{-1}$.

\subsection{Gaussian approximation for dual ensemble \\and the relation to DMFT}

Since the transformation from the initial system (\ref{Hubbard}) to
the action (\ref{Sf}) contains no approximations, it is equally hard
to describe exactly the properties of $c^*, c$ fermions as thereof
$f^*, f$ dual-fermions. The main idea of switching to the new
variables is that, for a properly chosen $\Delta$, correlation
properties of the $f^*, f$ system are simpler than for the $c^*, c$
original model. In other words, the magnitude of the nonlinear part
in the dual action can be effectively decreased by the proper choice
of $\Delta$. To illustrate this statement, let us just neglect $V$
in (\ref{Sf}). We denote the Green's function for such Gaussian
approximation for the dual potential with calligraphic letters. The
expression (\ref{Sf}) corresponds to
\begin{equation}\label{DMFTdual}
{\cal G}^{dual}_{\omega, k}=-g_\omega \left((\Delta_\omega -
\epsilon_k)^{-1}+g_\omega\right)^{-1} g_\omega.
\end{equation}
 Being combined
with the identity (\ref{exact}), this gives the formula
\begin{equation}\label{DMFT}
 {\cal G}_{\omega,
 k}=\left(g^{-1}_\omega+\Delta_\omega-\epsilon_k\right)^{-1}.
\end{equation}
One can recognize that this is exactly a DMFT expression for the
Green's function. Therefore we conclude that for a properly chosen
$\Delta$ already a Gaussian approximation for the dual potential
yields a reasonable result, as DMFT does. It is important to point
out that DMFT works well for the whole range of the parameters. In
contrast, the Gaussian approximation for the atomic limit of the
initial model (\ref{Hubbard}) makes no sense. In that aspect, the
dual potential $V$ is indeed smaller than $U$.

An argumentation can be presented to justify that the DMFT value of
$\Delta$ is a proper choice for the Gaussian approximation
(\ref{DMFTdual}, \ref{DMFT}). One of the reasons is described in the
Appendix B. It turns out that Feynman minimization criterion for the
Gaussian trial action, been formulated for the dual ensemble, gives
exactly the DMFT hybridization function. Another argument is
presented in the following subsection.

Once the dual potential is taken into account, it yields a
correction to the DMFT result. It is useful to introduce the dual
self-energy
\begin{equation}\label{Sigmadual}
    \Sigma_{dual} \equiv {\cal G}_{dual}^{-1}-G_{dual}^{-1};
\end{equation}
and the correction to the DMFT self-energy
\begin{equation}\label{Sigmaprime}
    \Sigma'\equiv {\cal G}^{-1}-G^{-1}.
\end{equation}
With these quantities, we can reexpress the exact relation
(\ref{exact}) in a particularly simple form
\begin{equation}\label{exactSigma}
    \Sigma'^{-1}_{\omega,k}=g_\omega+\left(\Sigma^{dual}_{\omega,k}\right)^{-1}.
\end{equation}
We note that this expression relates quite different quantities:
$\Sigma_{dual}$ and $\Sigma'$ characterize the corresponding lattice
problems and carry, in general, both momentum- and
frequency-dependence, whereas $g$ comes from the impurity model and
is local in space.

\subsection{Diagram series: general properties and the choice of hybridization function}

The main idea of our method is to consider a diagrammatic expansion
with respect to the dual potential $V$. We will later demonstrate
that already low-order diagrams of such a series bring an important
information about non-local correlations. The basic reasoning for
this is presented in the previous subsection: since the value of $V$
is in certain sense small, the first few terms of the perturbation
series with respect to $V$ can make sense. More detailed discussion
about the small parameters of the theory are presented in the next
sections; let us first present the general properties of the
diagrams under consideration.

\begin{figure}
\includegraphics[width=\columnwidth]{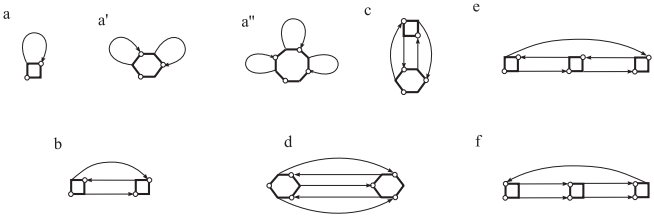}
\caption{Various diagrams for $\Sigma^{dual}$. Diagrams a, a$'$,
and a$''$ are vanished by the condition \ref{NoLoops}.}
\label{Diagram}
\end{figure}

The rules of diagram construction are quite similar to the usual
Matsubara diagram technique. The only difference from the standard
perturbation scheme is that the interaction operator $V$ is not
purely of the 4-th order form $f^* f^* f f$, and therefore
vertices in the diagrams are not necessarily four-leg, but may
formally have any even number of legs. For the choice
(\ref{definealpha}), these vertices are essentially
$\gamma^{(n)}$. They are connected with the lines being the dual
Green's functions. Some of the diagrams contributing yo the dual
self-energy are presented in Figure (\ref{Diagram}).

\begin{figure}
\includegraphics[width=0.5 \columnwidth]{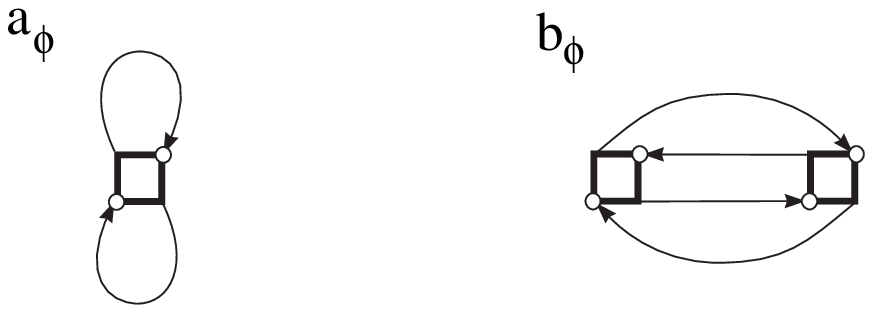}
\caption{Two simple diagrams for Baym functional
$\Phi_{dual}[G_{dual}]$. Functional differentiation of these
diagrams with respect to $G_{dual}$ produces diagrams $a$ and $b$
for self-energy.} \label{DiagramPhi}
\end{figure}

We use the skeleton diagrams with renormalized Green's functions,
so that the lines are complete $G_{dual}$, and not ${\cal
G}_{dual}$. The reason to use the skeleton-diagram expansion for
the dual self-energy is that it makes possible to obtain
conserving theories, similarly to conventional diagram technique
\cite{Baym}. The Baym criterion of a conservative theory is the
existence of a functional of the Green function $\Phi[G]$ such
that $\frac{\delta \Phi}{\delta G}=\Sigma$. Once this functional
is described by certain skeleton diagrams, taking the derivative
means just cutting the lines in that diagram. For example, the
diagrams (a) and (b) for the self-energy come from diagrams
(a$_\Phi$) and (b$_\Phi$), shown in Figure(\ref{DiagramPhi}) (of
course, care should be taken of the numerical factors).
Second-order differentiation with respect to $G$ gives the
two-particle quantities. Such a procedure automatically produces a
theory fulfilling the conservation laws for energy, momentum,
particle numbers etc.

In our consideration, the usage of skeleton diagrams describes a
corresponding Baym functional $\Phi_{dual}[G_{dual}]$ with the
functional derivative being $\Sigma_{dual}$. Therefore, it produces
a conservative approximation for the dual ensemble. Then it turns
out that the exact transformations (\ref{exact}) and
(\ref{exactGamma}) give a conserving description of the initial
system. Simply, the conservation laws imply certain selection rules
for $G$ and $\Gamma$, and (\ref{exact}, \ref{exactGamma}) clearly
preserve those selection rules during the transformation from dual
to initial quantities. More precisely, the conserving character of
an approximation in fact means that there exists some conserving
dual action $\tilde{S}[f, f^*]$, exactly corresponding to this
approximation. Since there is a one to one correspondence between
$S[f, f^*]$ and $S[c, c^*]$ (see the end of Section
\ref{ExactFormulas}), we conclude that the initial system described
by a certain $\tilde{S}[c, c^*]$ is also conserving.

Until now, the hybridization function $\Delta$ was formally not
specified. Now, we establish a condition for $\Delta$ that
corresponds to a particular condition for the diagrammatic series.
Let us again consider the DMFT. Suppose that we want to obtain the
DMFT result without DMFT loops, that is using $\Delta_\omega$ {\it
not fulfilling} (\ref{DMFTcond}). Formally, it is possible: one
should just sum up all the diagrams containing a single vertex
(diagrams a, a$'$, a$''$ etc.). Since these diagrams give exactly
the DMFT self-energy, such a procedure would indeed
recover the DMFT result for an arbitrary hybridization function. The
special DMFT choice of $\Delta$ just allows to eliminate such an
infinite summation, since (\ref{DMFTconddual}) eliminates all the
diagrams containing a simple closed loop. It is reasonable to keep
this property in higher approximations, that is to require
\begin{equation}\label{NoLoops}
        G^{dual}_{\omega,r=0}=0
\end{equation}
as a condition for $\Delta$. Then, all the diagrams with simple
closed loops drop out from the calculation. Note that these diagrams
however should be taken into account while taking the functional
derivatives. For example, the DMFT vertex part
$\Gamma_{dual}=\gamma^{(4)}$ comes out from the differentiation of
diagram (a). Finally, the condition (\ref{NoLoops}) obviously passes
into (\ref{DMFTconddual}) at the DMFT limit. Therefore, until the
corrections to DMFT are significant, one can approximate $\Delta_\omega$
with the DMFT hybridization function.

The vanishing of the closed loops seriously reduces the number of
the low-order dual diagrams. In most of the practical calculations
presented below we consider a single diagram (b). It is clear that
any reasonable expansion starts from this perturbation, and that
this diagram already incorporates some non-local physics. The
corresponding formula for the dual self-energy reads (spin and
orbital indices are omitted):
\begin{equation}\label{DiagramB}
    \Sigma^{dual}_{\omega, r}=\frac{1}{2 \beta^2}\sum_{\omega+\omega'=\omega_1+\omega_1}
    \gamma^{(4)}_{\omega \omega' \omega_1  \omega_2}
    \gamma^{(4)}_{\omega_2  \omega_1 \omega' \omega} G^{dual}_{\omega_1,
    r}G^{dual}_{\omega_2, r} G^{dual}_{\omega', -r}
\end{equation}

\subsection{Causal properties}\label{causalsection}

Beyond conservation laws, the Green's function should be causal.
The retarded Green's function $G^R(t)$, that is an analytical
continuation of $G_\tau$ to the real-time axis, should vanish for
negative time:
\begin{equation}\label{causal}
    G^{R}(t<0)=0.
\end{equation}

In the Fourier representation, condition (\ref{causal}) implies
the analyticity of $G_\omega$ in the upper complex plane, as this
follows directly from the definition of the Fourier transform. The
inverse is also true. If the Fourier transform of a function is
analytical in the upper-plane, the function is causal. To prove
this statement, it is enough to transform the integration contour
of the inverse Fourier transform away from the real axis.

Frequently, the causality principle is associated with the
positiveness of the imaginary part of the Green's function in the
real-frequency domain. For dual Green's function, this can lead to
certain misunderstanding. It is clear from condition
(\ref{NoLoops}) that the imaginary part of $G^{dual}$ cannot be
always-positive. However, this issue is purely formal. Condition
(\ref{causal}) itself does not imply that ${\rm Im} G_\omega$ is
positive. A trivial counter-example is the function $-G^{R}$. It
fulfills (\ref{causal}), and has an always-negative imaginary
part. We will argue the same for $G^{dual}$. It fulfills
(\ref{causal}). Although, its imaginary part is not
always-positive.

Let us illustrate this statement at the zeroth order of the
theory, single-site DMFT. It has been proven \cite{dmft,cdmft,dca}
that this theory is causal, so ${\cal G}$ and $g$ fulfill
(\ref{causal}). One can easily check, from the expressions
(\ref{DMFTdual}, \ref{DMFT}), for the case of DMFT,  a simple
relationship holds; ${\cal G}^{dual}={\cal G}-g$. It is
immediately clear from this formula, since ${\cal G}$ and $g$ are
causal, ${\cal G}^{dual}$ also fulfills (\ref{causal}). Note
again, both condition (\ref{NoLoops}) is fulfilled in DMFT, and
${-\rm Im} {\cal G}^{dual}_\omega$ is therefore essentially
non-positive.

Let us now consider the dual-fermion theory beyond DMFT. We will
show that, if the hybridization function $\Delta$ is casual, the
resulting Green's function is also causal. First of all, the
casuality of $\Delta$ is inherited from $g$ and $\gamma^{(n)}$.
Therefore, the dual system is characterized by the casual bare
propagator ${\cal G}^{dual}={\cal G}-g$ and casual interaction
operator. Therefore, the theory with skeleton diagrams results in
a causal $G^{dual}$ \cite{QFT}. Finally, it should just be proven
that the casuality $G^{dual}$ means the casuality of $G$. The
later statement follows from the exact relation (\ref{exact}).
Indeed, since $\alpha\equiv g_\omega$ does not have zeros in the
upper-plane, $g^{-1}_\omega$ is analytical. The same is true for
the quantity $(\Delta-\epsilon)^{-1}$. Therefore, the entire
right-hand side of (\ref{exact}) is analytical in the upper-plane.
This implies the causality of $G$.

In the calculation procedure described below, we always start from
a causal $\Delta$ and change it iteratively to deliver condition
(\ref{NoLoops}). We will argue that such an iteration procedure
preserve the causality of $\Delta$. Therefore, the entire theory
is causal.

Finally, let us recall  the issue of the positiveness of ${-\rm
Im} G$ in the complex upper-plane. Actually, this is related with
the positivity of the residuals, as it follows from the  Lehmann
representation $G=\sum \frac{Z_{mn}}{\omega-\omega_{mn}+i
\delta_{mn}}$. Here, the causality follows from the positivity of
$\delta$, whereas the requirement $Z>0$ ensures that ${-\rm Im}
G>0$. For our theory, we were not able to prove the positivity of
the residuals formally. However, we do not consider this as a
serious drawback, since our practical calculations always produce
undoubtly positive residuals.

\subsection{Small parameter in the extreme cases}

An important property of the DMFT approach is that it becomes exact
for the two opposite cases of a non-interacting Gaussian system and
of an extreme strong-coupling limit corresponding to the atomic
limit \cite{dmft}. The dual-fermion formalism inherits this
property; moreover the corresponding smallness appears in the
diagrams in a simple form. Let us first consider the strong-coupling
limit $\epsilon_k\to 0$. It is useful to estimate the DMFT dual
Green's function ${\cal G}$, defined by the formula (\ref{DMFTdual})
and the condition (\ref{NoLoops}). For a pure atomic limit
$\epsilon_k=0$, the Green's function is local, ${\cal G}_{r\neq
0}=0$. However, the local part of the Green's function also vanishes
due to the condition (\ref{NoLoops}). Formally, ${\cal G}\to 0$ as
$\Delta\to 0$. The  smallness of $\epsilon$ and $\Delta$ allows the
approximate estimation of the dual Green's function near the atomic
limit. It gives ${\cal G}_{\omega k}\approx g_\omega \epsilon_k
g_\omega$. Since the DMFT is almost exact near the atomic limit, the
same estimation is valid for $G^{dual}$. Consequently, {\it near the
atomic limit the lines in the dual diagrams carry a small factor}
$\epsilon_k$.

On the other hand, for the opposite weak-coupling limit $U\to 0$,
the vertex parts of the impurity problem can be estimated as
$\gamma^{(4)}\propto U, \gamma^{(6)}\propto U^2$, etc. Therefore,
{\it for the weak-coupling limit the vertices in the dual diagrams
are manifestly small}.

The presence of a small parameter in these two limits does not
guaranty a good interpolation between them. It should however be
mentioned, that the scheme performs well if the corrections to DMFT
are small: for this case we deal in fact with a perturbation series
around DMFT. The validity of the method for more general situations
should be checked in practical calculation. This practical validity
depends on the particular choice of diagrammatic approximation for
$\Sigma_{dual}$. In this context, it is worth to discuss the choice
of hybridization function $\Delta$.

\subsection{Calculation procedure}

\begin{figure}
\includegraphics[width=\columnwidth]{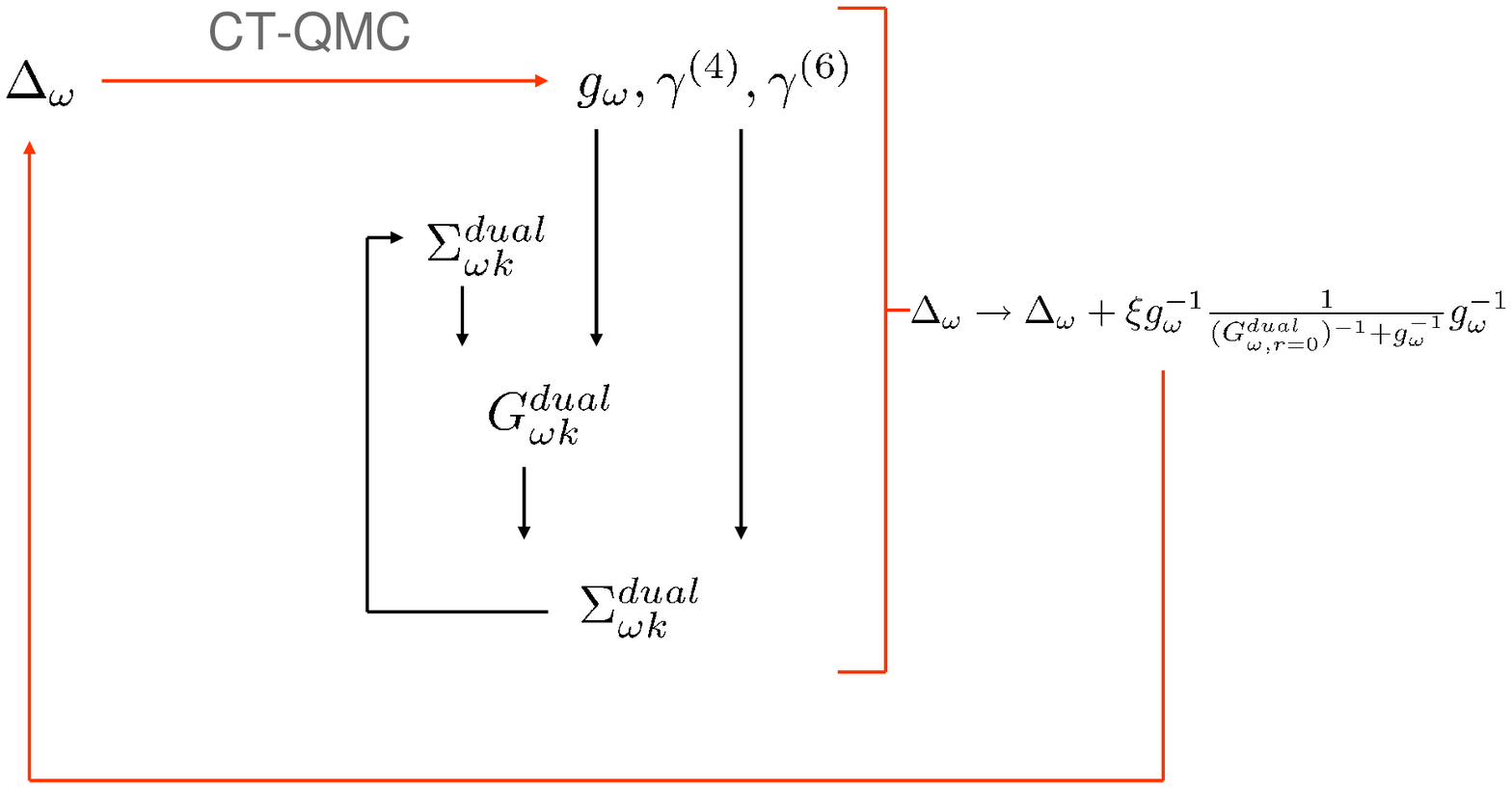}
\caption{(color online) The scheme of calculation. The calculation
includes ``big'' and ``small'' loop, marked with red and black
lines, respectively. The small loop is to determine the
renormalized dual Green's function $G^{dual}$ in a self-consistent
way, for given $\Delta, g$, and $\gamma^{(n)}$. The big loop is to
determine $\Delta$. Only the big loop requires a solution of the
impurity problem.} \label{Scheme}
\end{figure}

In practical calculations the solution was obtained iteratively,
similarly to the DMFT loop. The iterative scheme is presented in
Figure \ref{Scheme}. It includes the big (outer) and small (inner)
loops. The small loop is devoted to obtain the dual Green's function
and self-energy, given the solution of the impurity model with
certain $\Delta$. It starts from some guess for $\Sigma_{dual}$, for
instance $\Sigma_{dual}^{(0)}=0$. The dual Green's function $({\cal
G}_{dual}^{-1}-\Sigma_{dual})^{-1}$ is substituted in formula
(\ref{DiagramB}) to produce a new estimation for $\Sigma_{dual}$.
The procedure is repeated until converging results are reached.

The big loop is very similar to the DMFT iterative procedure. We
start with some initial guess for $\Delta$ and solve an impurity
model. We use the weak-coupling CT-QMC solver \cite{Rub05}, which
produces both the Green's function $g$ and the 4-point vertex
$\gamma^{(4)}$ in the frequency domain. Then we perform the inner
loop to obtain $G_{dual}$ (this step is not necessary in DMFT, since
it uses the bare dual Green's function ${\cal G}_{dual}$). Finally,
we take a new guess for the hybridization function
\begin{equation}\label{NewDelta}
    \Delta_\omega \to \Delta_\omega+\xi ~g_{\omega}^{-1}\frac{1}{(G^{dual}_{\omega, r=0})^{-1}+g_{\omega}^{-1}}g_{\omega}^{-1}
\end{equation}
and repeat the self-consistent procedure. A value of the parameter
$\xi \leq 1$ was chosen to ensure better convergence.  The last
formula is organized in such a way that (i) its fixed point
clearly satisfies the condition (\ref{NoLoops}) and (ii) for
$\Sigma_{dual}=0$ it passes into the DMFT update formula
$\Delta_{\omega}\to\Delta_{\omega}+\xi ({\cal G}_{\omega,
r=0}^{-1}-g_\omega)$. Of course, only the requirement (i) is
actually necessary, so that formula (\ref{NewDelta}) is not
unique. In particular, it is useful to consider an update
\begin{equation}\label{NewDelta1}
    \Delta_\omega \to \Delta_\omega+\xi ~g_{\omega}^{-1} G^{dual}_{\omega, r=0}
    g_{\omega}^{-1}.
\end{equation}
One can easily see that an update (\ref{NewDelta1}) conserves
causal properties of $\Delta$, do that the convergence of the
iteration process (\ref{NewDelta1}) proves the causality of the
result. Such a convergence indeed takes the place for the
calculations presented below. Note also that near the fixed point
$G^{dual}_{\omega, r=0}=0$ formula (\ref{NewDelta}) passes into
(\ref{NewDelta1}), so that there is no much practical difference
between these two formulas.

\section{Application to the Hubbard model}

In the next sections, we present the results of our calculations for
the 2D Hubbard model. We start with the half-filled case with
next-nearest neighbor hopping $t'=0$ lattice. We compare our data
with direct QMC simulations on a finite Hubbard lattice, which are
relatively simple due to the absence of the sign problem for the
half-filled Hubbard model.

Properties of the half-filled Hubbard model are well-known and are
mostly related to the antiferromagnetic phenomenon and Mott
metal-insulator transition. Local magnetic moment on atoms are
formed and tend to ordered into an antiferromagnetic lattice due to
the effective super-exchange coupling. At zero temperature, the
antiferromagnetism arises already at $U=0^+$, because of the perfect
nesting. At finite temperature, the true antiferromagnetism is
destroyed by the long-range fluctuations. However, short-range
antiferromagnetic correlations are still present. Short-range
antiferromagnetic ordering manifests itself as the strong pseudogap
in the local electron spectral function.

We consider the system with $t=0.25$ at inverse temperature
$\beta=20$, with different values of $U$. Since the temperature is
relatively high, it is enough to use the reference data obtained
just for the $8 \times 8$ lattice QMC simulation, with subsequent
maximum-entropy continuation of the data to obtain local density of
states (DOS). The result for paramagnetic calculation is presented
in Figs. \ref{FigGunpolar}, \ref{FigGunpolar10} (thin solid line).
These results show that the narrow antiferromagnetic pseudogap is
formed at approximately $U=1.0$. For larger $U$, the DOS contains
also a wider Mott gap, having a halfwidth of about $U/2$. At
$U=2.0$, the system shows essentially Mott-insulator DOS; the effect
of antiferromagnetism in this case consists in the sharp shoulders
of the Mott gap.

To understand better the physics of the half-filled Hubbard model,
it is worth to analyse the behavior of the electronic self-energy
$\Sigma$. At small $U$, this is a small regular correction to the
dispersion law $\epsilon_k$. It follows from the weak-coupling
analysis that ${\rm Im} \Sigma$ is strongly anisotropic in this
regime, with peaks near $[0,\pm \pi], [\pm \pi,0]$ points. In
contrast, for the truly antiferromagnetic gap, $\Sigma_k$ would have
a pole at the Fermi surface. The residue of this pole is the same at
all points of the Fermi surface. For large enough $U$ this pole is
somehow shifted from the real-frequency axis due to long-range
thermal fluctuations, but the qualitative picture remains the same:
a sharp peak in ${\rm Im} \Sigma$, with almost constant magnitude
along the Fermi surface.

It is well-known that doping changes the physics of the Hubbard
model substantially. First of all, already a few-percent doping
suppresses the antiferromagnetism. At higher doping values there is
a trend to $d$-wave superconductivity. 
A superconducting phase has been obtained in various cluster-DMFT calculations
\cite{cdmftdw,jarrell1} near the optimal doping of about 15\%. This agrees well with the phase
diagram of high-$T_c$ cuprates \cite{dwave}. The pseudogap formation in the doped Hubbard model was first analyzed by the
cluster DMFT method (more specifically, Dynamical Cluster Approximation) in Ref.
\onlinecite{jarrell2}. For further applications of the DCA to the 2D Hubbard model,
see Refs. \onlinecite{dca,jarrell3, jarrell4, jarrell5}. In the following
consideration, we will not discuss the superconductivity itself, but
we will address the so-called Fermi arc phenomenon. Essentially,
this is an anisotropic destruction of the Fermi surface in the
pseudogap regime. Only the parts of Fermi surface near the nodal
direction remain well-defined at low temperature. In the anti-nodal
direction, the spectral function at the Fermi level is vanishingly
small.

A methodological difference between the doped and the undoped cases
is that the sign problem makes direct lattice simulations away from
half filling practically impossible \cite{signQMC}. Therefore, the
reference point can only be the results of different approximate
schemes or the experimental data.

\subsection{Undoped case: translationally-invariant solution}

First, we discuss the result of the dual fermion investigation
without a spontaneous symmetry breaking which means that the
impurity problem is assumed to have no spin-polarization. The data
presented in this chapter have been partly discussed previously as a
Brief Report \cite{Brief}.

The translationally-invariant DMFT predicts a Mott transition at
rather high value $U>3.0$ (for a bandwidth $W=8 t=2.0$). It is
important to point out that the density of states at the Fermi
energy is independent of $U$ within the entire Fermi-liquid phase.
This is a consequence of the locality of the self-energy in DMFT.
Therefore, for $U\approx 1.5-3.0$, the approximation predicts a
three-peak DOS which consists on two Hubbard bands at $\pm U/2$ and
a Kondo-like central peak providing the `pinned' value of DOS at
Fermi level.

\begin{figure}
\begin{center}
\includegraphics[width=0.9 \columnwidth]{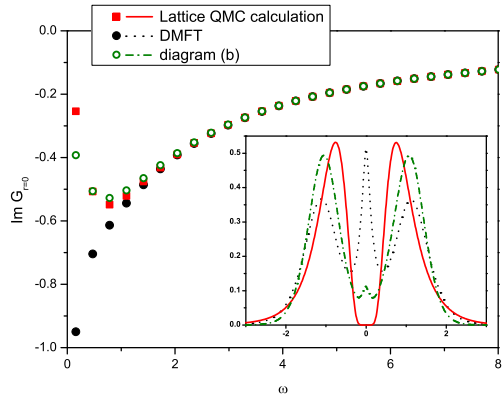}
\end{center}
\caption{(color online) Local Green's function at Matsubara
frequencies and density of states for undoped Hubbard model at
$t=0.25, U=2.0, \beta=20$. The results of DMFT and the calculation
with nonlocal diagram correction (b) are compared with the
reference data obtained for  $8\times 8$ lattice QMC simulation.}
\label{FigGunpolar}
\end{figure}

This behaviour is inconsistent with the reference data described
above. Actually, those data do not show a three-peak structure,
because of the antiferromagnetic pseudogap. Besides
antiferromagnetism, the DMFT sufficiently overestimates the
critical value of $U$ for the Mott transition: according to the
reference data, the the system shows DOS of the Mott-insulator
nature already at $U\approx 2.0$ (see Figure \ref{FigGunpolar}).

\begin{figure}
\includegraphics[width=\columnwidth]{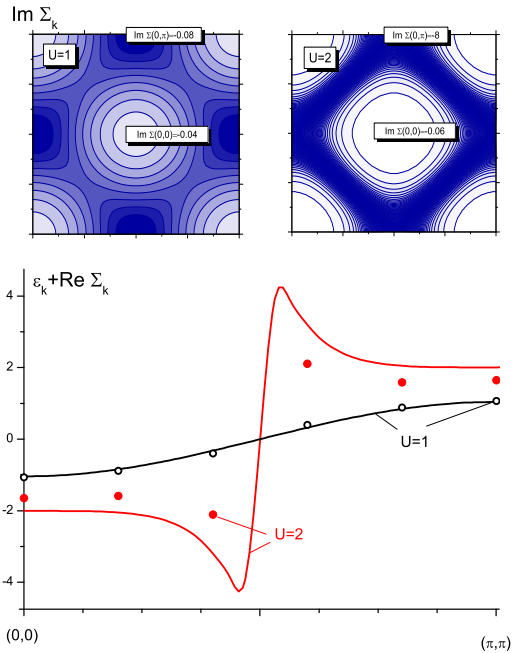}
\caption{(color online) Momentum dependence for the self-energy
function at Fermi energy, obtained with diagram (b) within the
translationally-invariant approximation for the undoped Hubbard
model. Data are shown  at $t=0.25, \beta=20$, for $U=1.0$ and
$U=2.0$. Upper panel: contour plots for $k$-dependence of the
imaginary part of the self energy. Lower panel: renormalized
dispersion law  $\epsilon_k+{\rm Re} \Sigma_{\omega=0, k}$,
compared with the reference data obtained for  $10\times 10$
lattice. The Figure has been published 
previously in the Breief Report \onlinecite{Brief}.} \label{FigDispunpolar}
\end{figure}

Let us  take the leading dual diagram (b) into account. The
corresponding data is presented in Figs.  \ref{FigGunpolar},
\ref{FigDispunpolar}. Since the self-energy is not local anymore,
there is no pinning at Fermi level, and the Kondo-like peak
disappears. Furthermore, the self-energy momentum dependence agrees
well with the qualitative picture described above. The upper panel
of Figure  \ref{FigDispunpolar}  presents contour plots for ${\rm
Im} \Sigma_{\omega=0, k}$ at $U=1.0$ and $U=2.0$ (the data are
obtained by a polynomial extrapolation from the Matsubara
frequencies). The value of ${\rm Im} \Sigma_{\omega=0, k}$ grows
dramatically as $U$ changes from $1.0$ to $2.0$. At larger $U$,
there is an expected sharp  non-Fermi liquid peak in ${\rm Im}
\Sigma_{\omega=0, k}$ at Fermi level, without a remarkable
anisotropy along Fermi surface. At smaller $U$, the peak is
broadened, with maxima near van Hove singularities. The renormalized
dispersion law $\epsilon_k+{\rm Re} \Sigma_{\omega=0, k}$ is now
also in a qualitative agreement with numerical data, as the lower
panel of Figure \ref{FigDispunpolar}. In these graphs,
$\epsilon_k+{\rm Re} \Sigma_{\omega=0, k}$ is compared with the
reference data for a $10\times 10$ lattice. There is a qualitative
difference between the results for $U=1.0$ and $U=2.0$: for later
case the corrections are quite large so that there is a dependence
resembling $\epsilon_k^{-1}$. The superiority of the result against
DMFT should be stressed, as there is no $k$-dependence of $\Sigma$
in the DMFT approach.

\begin{figure}
\begin{center}
\includegraphics[width=0.9 \columnwidth]{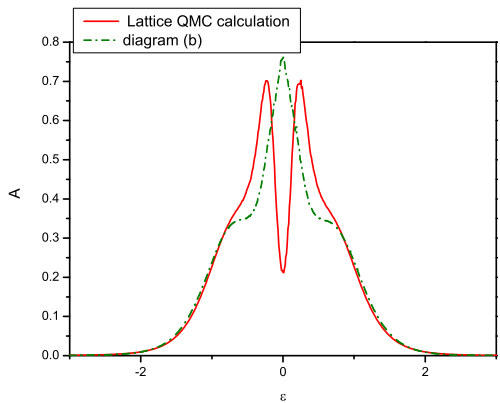}
\end{center}
\caption{(color online) Density of states for undopped Hubbard
model at $t=0.25, U=1.0, \beta=20$. The result of the
translationally-invariant calculation with diagram (b) is compared
with the reference data for $8\times 8$ lattice. An
antiferromagnetic pseudogap is pronounced in the reference data
and does not appear in the approximation.} \label{FigGunpolar10}
\end{figure}

Let us point out the drawbacks of the present results. First of
all, there is still no perfect quantitative agreement with the
reference data, although the DMFT result is improved remarkably.
The source of this discrepancy becomes clear when the DOS for
$U=1.0$ is plotted (Figure \ref{FigGunpolar10}). The pseudogap is
much narrower in this case. It resembles the situation for $U\to
+0$ at zero temperature, then an antiferromagnetic ordering appears
due to long-range nesting phenomena. Its evident from Figure
\ref{FigGunpolar10} that the calculation with dual diagram (b)
does not reproduce this pseudogap at all. Back to the results for
$U=2.0$, the pseudogap in our calculation appears to be not as
deep and not as steep, as it should be (Figure \ref{FigGunpolar}).
We have tried to take higher diagrams into account and found out
that it does not help much. We conclude that the dual-fermion
corrections, as they considered above, improve the description of
short-range Mott physics, but they do not take the long-range
antiferromagnetic fluctuations into account.

To explain this failure, let us recall the Hubbard model with small
$U$ at zero temperature. As pointed above, our technique passes into
weak-coupling diagram expansion for $U\to 0$. But it is clear that
the weak-coupling expansion is suitable for the metallic phase only
and cannot reproduce the antiferromagnetism, since this is a
non-perturbative phenomenon \cite{AFM}.  Evidently, the dual-fermion
expansion inherits this property. The best possible achievement
within this framework would be to obtain a phase transition, where
the corresponding susceptibility diverges \cite{dualsusc}.

There are two ways to take antiferromagnetism into account. First,
one can switch from single-site to cluster DMFT. Thus the
antiferromagnetic phase transition maintains the periodical symmetry
of the superlattice made of clusters, there is no problem with
non-analyticity in this case. Indeed, various cluster DMFT
approaches \cite{dca,cdmft} reproduce the antiferromagnetic gap. The
dual-fermion corrections can be used to improve the accuracy of
those methods \cite{dualsusc,Haf07}.

The second option is to stay with the single-site starting point,
but allow for the antiferromagnetic ordering on the lattice. In this
case, the effective impurity problem is spin-polarized. It is known
that such an approach indeed works quite well already at the DMFT
level \cite{DMFTafm}. It can be expected, that the dual-fermion
technique can effectively provide the correction in this case. The
next section describes such a theory and the corresponding results.

\subsection{Undoped case: antiferromagnetic symmetry breaking}

For clarity, let us present the explicit expressions for this
case. The antiferromagnetism means that the primitive cell is
doubled. The dual Green's function, as well as other
single-electron quantities of the antiferromagnetic state, depends
on the difference of the two coordinate arguments and single spin:
$G^{dual}_{\omega, j, s, j', s'}=G^{dual}_{\omega, j-j', s}$ (note
that $s'$ is defined by $s$ and $r=j-j'$). Given
$G^{dual}_{\omega, r, s}$, it is easy to obtain
$\Sigma^{dual}_{\omega, r, s}$ from the formula (\ref{DiagramB}).
In this expression, the spin dependence of $\Sigma$ comes from the
spin polarization of $G^{dual}$ and, in principle, of the
vertex $\gamma^{(4)}$. However, the numerical result
for the latter quantity appears to be quite noisy. Therefore, we
neglected the spin polarization of $\gamma^{(4)}$, performed a
averaging over spin orientation, and thus operated with the tensor
of the `paramagnetic' symmetry. Such a tensor has the two
independent components $\gamma'\equiv \gamma^{(4)}_{ssss}$ and
$\gamma''\equiv \gamma^{(4)}_{ss-s-s}$, so that the expression
(\ref{DiagramB}) becomes
\begin{widetext}
\begin{equation}\label{DiagramBAF}
    \Sigma^{dual}_{\omega, r,s}=\frac{1}{2 \beta^2}\sum_{\omega+\omega'=\omega_1+\omega_1}
    \gamma'_{\omega \omega_1 \omega' \omega_2}
    \gamma'_{\omega_2 \omega' \omega_1 \omega} G^{dual}_{\omega_1,
    r,s}G^{dual}_{\omega_2, r,s} G^{dual}_{\omega', -r,s}+
\frac{1}{\beta^2}\sum_{\omega+\omega'=\omega_1+\omega_1}
    \gamma''_
    {\omega \omega_1 \omega' \omega_2}
    \gamma''_{\omega_2 \omega' \omega_1 \omega} G^{dual}_{\omega_1,
    r}G^{dual}_{\omega_2, r, -s} G^{dual}_{\omega', -r, -s}
\end{equation}
\end{widetext}
We believe that this approximation is valid, since the most
important contribution to the  symmetry-break arises from the
spin-polarization of the single electron quantities $g, \Delta$,
and $\Sigma^{dual}$, entering the expression for $G^{dual}$.

The next step is to write explicitly the definition
$\Sigma_{dual}={\cal G}_{dual}^{-1}-G_{dual}^{-1}$ in the momentum
space. Here,  the $2\times 2$ matrices must
 be used, as the momentum is conserved up to $Q=(\pi, \pi)$.
 Let us denote $G^{dual
(0)}_{\omega, j-j'}=\frac{1}{2}(G_{\omega, j-j', s}+G_{\omega,
j-j', -s})$ and $G^{dual (AF)}_{\omega,
j-j'}=\frac{1}{2}(G_{\omega, j-j', s}-G_{\omega, j-j', -s})$. It
is easy to check that the definition $\Sigma_{dual}={\cal
G}_{dual}^{-1}-G_{dual}^{-1}$ stays fulfilled with the matrix
$$\left(%
\begin{array}{cc}
  G^{dual (0)}_k & G^{dual (AF)}_k \\
  G^{dual (AF)}_k & G^{dual (0)}_{k+Q} \\
\end{array}%
\right)$$ used for $G^{dual}$, and similarly for  $\Sigma_{dual},
{\cal G}_{dual}$. This gives a way to construct $G^{dual}$ from a
given $\Sigma^{dual}$  and thus close the inner iteration loop. The
self-consistency condition (\ref{NoLoops}) remains unchanged, so
that the big loop is essentially the same. Finally, the exact
relationship (\ref{exact}) can be written in the matrix form, giving
thus a complete description of the antiferromagnetic state. Of
course, the same treatment with $\Sigma^{dual}=0$ corresponds to the
antiferromagnetic DMFT.

\begin{figure}
\includegraphics[width=\columnwidth]{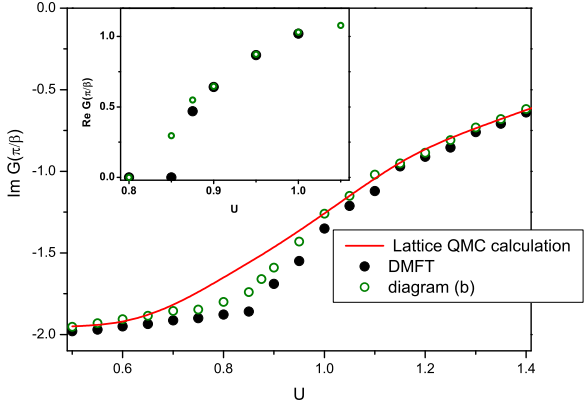}
\caption{(color online) Results of DMFT calculation and the scheme
with diagram (b), taking the antiferromagnetic ordering into
account. The results for local part of the Green's function at
lowest Matsubara frequency are compared with reference data for
undoped Hubbard model at $t=0.25, \beta=20$. QMC calculation at
$8\times 8$ lattice are used for reference.} \label{FigScanUpolar}
\end{figure}

Actually, once the antiferromagnetism is taken into account, the
DMFT result itself already is not too bad. The corresponding data
are presented in Figure \ref{FigScanUpolar}, where we show how the
Green's function at the
lowest Matsubara frequency depends on $U$. 
At small U, the system is a normal Fermi-liquid. There are small
corrections due to the correlations. Of course, DMFT cannot
reproduce the anisotropy of the self-energy, but the description of
local Green's function is pretty good. For large $U$, the system
exhibits a strong antiferromagnetism, which is destroyed only at
long-range scale. In DMFT, the antiferromagnetic ordering appears in
this range. The simplest way to take the long-range fluctuations
into account within DMFT framework is to average over the two
antiferromagnet sub-lattices. This eliminates the real part of the
Green's function. A comparison of ${\rm Im} G_{\pi/\beta,r=0}$ with
lattice QMC simulations again shows a good agreement (the
antiferromagnetic regime starts from $U\approx0.85$, as the inset in
Fig.\ref{FigScanUpolar} shows).
The largest deviations of the DMFT result from the reference data
occur in the intermediate regime $U\approx 1$. Probably, in this
regime the fluctuations are essentially non-local but still
mid-range. Therefore they cannot be described as a static
long-range antiferromagnetic ordering.

\begin{figure}
\begin{center}
\includegraphics[width=0.9 \columnwidth]{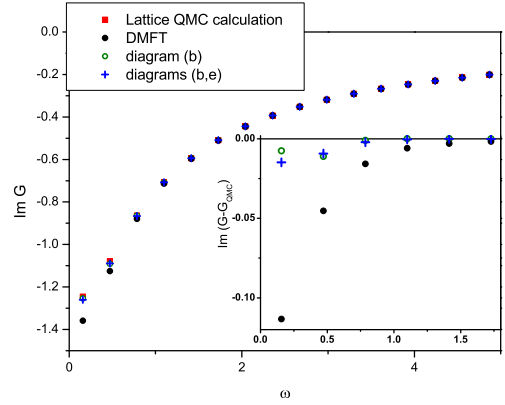}
\end{center}
\caption{(color online) Imaginary part of the local Green's
function of undoped Hubbard model at Matsubara frequencies. The
data are shown for $U=1,~t=0.25, ~\beta=20$. The reference data
are compared with the results of approximate schemes taking
antiferromagnetism into account. The results of DMFT calculation,
of the scheme with diagram (b), and of the approximation taking
two diagrams (b), (e) into account are shown. Inset shows the
deviation of the approximate results from reference data.}
\label{FigDOSLadderG}
\end{figure}

\begin{figure}
\begin{center}
\includegraphics[width=0.9 \columnwidth]{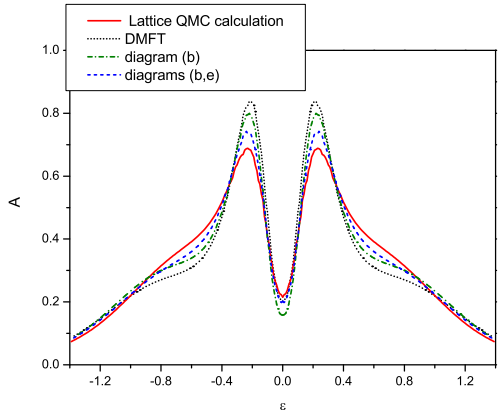}
\end{center}
\caption{(color online) Density of states of the undoped Hubbard
model, restored from the data presented in Figure
\ref{FigDOSLadderG}. The approximate result becomes closer to
reference data as diagrams (b) and (e) are taken into account.}
\label{FigDOSLadder}
\end{figure}

The same Fig. \ref{FigScanUpolar}  presents the result obtained with
the first nonlocal dual diagram (b). In this calculation, we again
allow for the antiferromagnetism. The symmetry breaking down a
almost the same value of $U$, and the magnetization coincides the
DMFT result. There is however a remarkable correction to ${\rm Im}
G_{\pi/\beta,r=0}$. Near both limiting cases, the reference
dependence is reproduced very well, since the diagram (b) yields a
leading-order correction to the already good DMFT result. In the
`critical' intermediate regime, the situation is not as good.
However, the correction still behaves regularly and shows the
correct trend. It is also important that while the DMFT data for
${\rm Im} G_{\pi/\beta,r=0}$ show a clear kink at the transition
point, the dual-diagram correction makes the curve much smoother.
This is certainly more physical, because the reference lattice QMC
data contains no singularities, since there is no true
phase-transition.

We did not found that any particular higher-order diagram improves
the result for $G_{\pi/\beta,r=0}$ significantly. This indicates
that a large number of higher-order diagrams contribute the
result. Actually, this is an expectable situation near the
critical point. However, it was found that higher-order ladder
corrections give a particularly important contribution to the
spectral function of the system. Let us illustrate this statement,
using the data for $U=1.0$. The Green's function at Matsubara
frequencies for this case are plotted in Figure
\ref{FigDOSLadderG}. Since the points with dual-diagram
corrections are very close to the reference ones and can hardly be
distinguished, we plot also the difference from the reference
lattice QMC result in the inset of Figure \ref{FigDOSLadderG}.
Figure \ref{FigDOSLadder} shows the maximum-entropy guess for the
corresponding DOS. Since the problem of analytical continuation of
the Green's function to the real-frequency axis is known to be
ill-posed, we took special measures while calculating the density
of states. The Green's functions are computed with high accuracy,
and the maximum-entropy analytical continuation is performed with
the same a priory parameters for all curves. This ensures that the
graphs for the spectral function can be compared one with another.
The spectral function clearly illustrates what is the physical
origin of the discrepancy between the DMFT and reference data.
Indeed, since DMFT replaces the nonlocal dynamical
antiferromagnetic correlations with static ordering, it
overestimates the antiferromagnetism in the model. Therefore the
pseudogap appears to be too deep; its shoulders and Hubbard bands
in the DMFT graph are narrower than they should be. The situation
is partly improved for the diagram (b): the shoulders and Hubbard
bands are closer to the reference curve although the estimation at
Fermi energy looks worse. The serious improvement arises from the
next diagram of the ladder, as the dash-dot curve in Figure
\ref{FigDOSLadder} shows. This is very expectable, because the
long-range antiferromagnetic fluctuations are exactly described by
these ladders.On the other hand it is interesting to observe from
the inset in Figure \ref{FigDOSLadder} that this diagram does not
improve the result for $G_{\pi/\beta,r=0}$, but makes its
deviation from the reference data more regular.

\subsection{Doped Hubbard: Fermi arcs formation and flattening of the dispersion
law}\label{DopedSection}

Here we present the results obtained with the dual-fermion technique
for the pseudogap regime, which corresponds to the doping below
optimal and relatively high temperature. We use the
rotationally-invariant approximation, so the effects of
superconductivity and antiferromagnetism were not included in the
theory. However it turns out that the theory still captures the
physics responsible for the Fermi arc formation, and yields results
which compare well to experimental data.

To make the simulation more realistic we introduce the next-neighbor
hopping term $t'$. The parameters of the model are $U=4.0, t=0.25,
t'=-0.075, \beta=80$. The ratio $t'/t \approx - 0.3$ roughly
corresponds to the case of YBa$_2$Cu$_3$O$_7$ \cite{YBCO}. The
relatively large value of $U=2W$ was taken because there is
experimental evidence that the system should be a Mott insulator at
small doping, which requires $U>1.5 W\approx 3.0$. The temperature
used roughly corresponds to 100-150 K, which is a proper value for
the pseudogap phenomena in high-temperature superconducting
materials. Most of the results are presented on doping level of 14\%
.

\begin{figure}
\begin{center}
\includegraphics[width=0.9 \columnwidth]{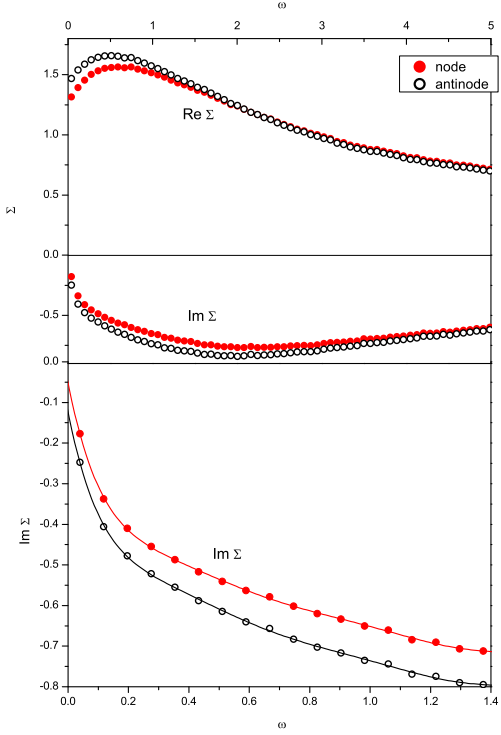}
\end{center} \caption{Self-energy function of
$tt'$ Hubbard model $\Sigma_{\omega, k}$ at nodal and antinodal
points of the Fermi surface at Matsubara frequencies. Diagram (b)
is used for the calculations. The data are plotted for 14 \%
doping $tt'$ Hubbard model at $t=0.25,~ t'=-0.075,~ U=4.0,
\beta=80$. Upper panel: real and imaginary parts of
$\Sigma_{\omega, k}$. Lower panel: ${\rm Im} \Sigma_{\omega, k}$
in a low-frequency region and its approximation with a 7-th order
polynomial.} \label{FigSigmaNAN}
\end{figure}

\begin{figure}
\begin{center}
\includegraphics[width=0.9 \columnwidth]{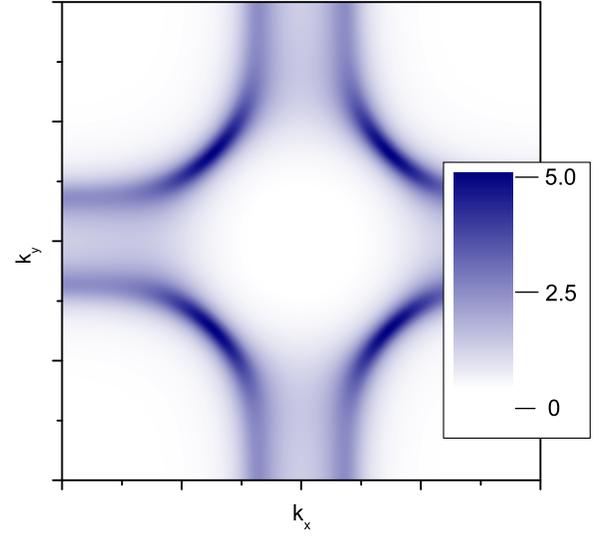}
\end{center} \caption{Spectral function $A_{\omega=0,
k}$ at Fermi level: the calculation with diagram (b) and
polynomial extrapolation from Matsubara frequencies. Parameters of
the Hubbard model are the same as in Figure \ref{FigSigmaNAN}. }
\label{FigSpectralFunctionDoped}
\end{figure}

Figure \ref{FigSigmaNAN} presents the results obtained for the
self-energy $\Sigma_{\omega, k}$ at the nodal and anti-nodal points
of the Fermi surface. The position of Fermi surface was defined as a
maximum of the spectral density. A polynomial extrapolation for
$\Sigma_{\omega}$ was constructed to obtain the imaginary part of
self-energy at Fermi level. One can observe a remarkable difference
in the low-energy limit of $ \Sigma_{\omega, k}$ at the nodal and
anti-nodal points: the corresponding values of ${\rm Im}
\Sigma_{\omega=0, k}$ differ approximately by a factor of two. The
spectral function $A_k=(2 \pi)^{-1} {\rm Im} G_{\omega=0, k}$ for
the entire Brillouin zone is mapped in Figure
\ref{FigSpectralFunctionDoped} for 14\% doping. The Fermi surface in
the antinodal direction is quite diffuse, in accordance with the
experimental results.

\begin{figure}
\begin{center}
\includegraphics[width=0.9 \columnwidth]{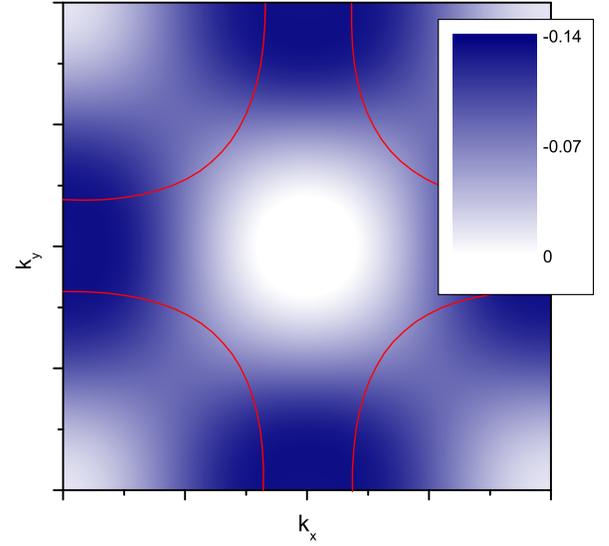}
\end{center}\caption{Imaginary part of the self energy
 ${\rm Im}
\Sigma_{\omega=0, k}$ at Fermi level: the calculation with diagram
(b) and polynomial extrapolation from Matsubara frequencies.
Parameters of the Hubbard model are the same as in Figure
\ref{FigSigmaNAN}. The red line indicates Fermi surface.}
\label{FigSigmaKDoped}
\end{figure}

\begin{figure}
\begin{center}
\includegraphics[width=0.9 \columnwidth]{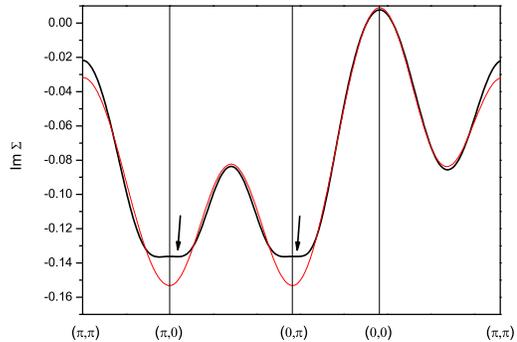}
\end{center}\caption{Imaginary part of the self energy
 ${\rm Im}
\Sigma_{\omega=0, k}$ at Fermi level: the calculation with diagram
(b) and polynomial extrapolation from Matsubara frequencies. Solid
line shows the same data as presented in Figure
\ref{FigSigmaKDoped}. Dot line is a fit with the next-neighbor
Fourier components. Arrows mark the flattened region at the
antinodal direction. Positive sign of ${\rm Im} \Sigma_{\omega=0,
k}$ in a small region near the $(0,0)$ is probably an artifact of
the polynomial extrapolation procedure.} \label{FigCutSigma}
\end{figure}

It is worth to consider the spatial dispersion of the self-energy
function. The map of ${\rm Im} \Sigma_{\omega=0, k}$ is presented
in Figure \ref{FigSigmaKDoped}, whereas Fig. \ref{FigCutSigma}
shows the behavior of this quantity along the
$(\pi,\pi)-(\pi,0)-(0,0)-(\pi,\pi)$ contour. The data are obtained
with a polynomial extrapolation from Matsubara axis. The estimated
errorbar of the extrapolation procedure is 0.01. An interesting
property of the data obtained is that $\Sigma_{\omega=0, k}$
appears to be substantially non-local, but still short-range.
Actually, the data of Figs. \ref{FigSigmaKDoped},
\ref{FigCutSigma} can be approximately described by the
next-neighbor approximation, that is, the most important
components of $\Sigma_{\omega=0, R}$ are $\Sigma_{R=(0,0)},
\Sigma_{R=(0,1)}$ and $\Sigma_{R=(1,1)}$. The doted line in Fig.
\ref{FigCutSigma} is produced with these Fourier-components only,
and it is quite consistent with the initial curve, except  the
points $(\pi,0)$ and $(0,\pi)$ where the self-energy is flattened.
It is worth to notice also that ${\rm Im} \Sigma_{\omega=0, k}$ is
maximal there. Interestingly, variational cluster calculations
\cite{Hanke} demonstrate that near the nodal point, in contrast
with the antinodal one, the superconducting gap (that is,
anomalous part of the self energy) also can be described in the
nearest-neighbor approximation.

\begin{figure}
\begin{center}
\includegraphics[width=0.9 \columnwidth]{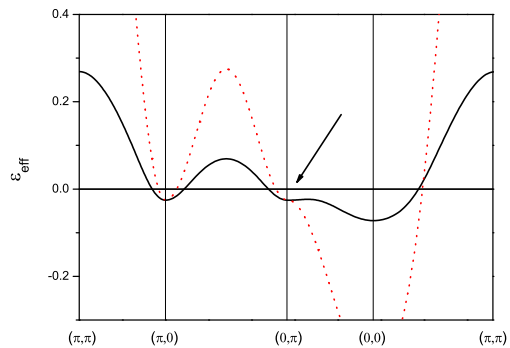}
\end{center}\caption{Quasiparticle dispersion law, defined from formula
(\ref{Dispeff}) (thick line), compared with initial dispersion
(thin line). Model parameters are the same as in Figures
\ref{FigSigmaNAN}-\ref{FigCutSigma}. Arrows mark the flattening of
the van Hove singularity.} \label{FigCutDisp}
\end{figure}

Figure \ref{FigCutDisp} shows the effective quasiparticle energy,
defined by the formula
\begin{equation}\label{Dispeff}
   \epsilon^{eff}_k={\rm Re}\left[\frac{\epsilon_k-\mu+\Sigma_{\omega=0,k}}{1+i
   \frac{\partial}{\partial \omega}\Sigma_{k,\omega=0}}\right].
\end{equation}
The initial dispersion law $\epsilon_k$ is shown in the same Figure
with thin line. One can see an narrowing of the quasiparticle band,
mainly due to the $\partial \Sigma_{\omega, k} /
\partial \omega$ term. The latter is large due to a closeness to the
Mott transition point. Another important change is again the
flattening of the curve near $(0,\pi)$ point.

A flattening of the dispersion curve near the antinodal point was
earlier predicted \cite{VH,vanHove} as due to a non-Fermi-liquid
behavior when the Fermi energy crosses  van Hove singularity. The
main conclusion of \cite{VH,vanHove} is that in the
strong-interacting regime van Hove point expands to a finite region
of the Fermi surface, where the dispersion law is flattened. The
$k$-dependence of the self-energy and vertex function are of crucial
importance for this phenomenon.

Its worth to note that cluster calculation hardly can reproduce
the result for the van Hove behavior, because the flattened region
is much smaller that the entire Brillouin zone.

\begin{figure}
\begin{center}
\includegraphics[width=0.9 \columnwidth]{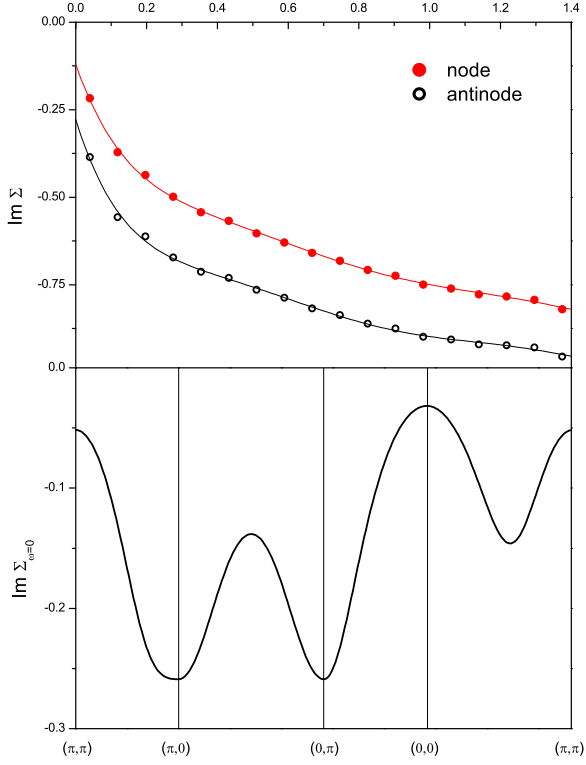}
\end{center}\caption{Imaginary part of the self-energy function for the 7\% doped system.
Other parameters of the $tt'$ Hubbard model are the same as in
Figures \ref{FigSigmaNAN}, \ref{FigCutSigma}. Upper panel: ${\rm
Im} \Sigma$ at the nodal and antinodal points of the Fermi
surface, and its polynomial fit at Matsubara frequencies. Lower
panel: low-energy behavior of ${\rm Im} \Sigma$ at the
$(\pi,\pi)-(\pi,0)-(0,0)-(\pi,\pi)$ contour.} \label{Fig7perc}
\end{figure}

We also performed calculations for other doping. Figure
\ref{Fig7perc} is devoted to ${\rm Im} \Sigma$ at 7\% doping.
Smaller doping makes the system closer to Mott insulator, therefore
the value of ${\rm Im} \Sigma$ is substantially larger then for the
14\% doped system (Figs. \ref{FigSigmaNAN}, \ref{FigCutSigma}). The
flattened regions disappear in this case. However, there is still a
clear difference between the nodal and antinodal directions in the
low energy limit: the  values of ${\rm Im} \Sigma_{\omega=0}$ at
these points differ by a factor of two.

Finally, a few words should be said about the region near $(0, 0)$
point in Figure \ref{FigCutSigma}, where our polynomial fit
predicted slightly positive ${\rm Im} \Sigma$ (that corresponds to
${\rm Im} G<0$). We argue here that this is merely an artifact of
the extrapolation procedure. Indeed, as it is discussed in section
\ref{causalsection}, negative ${\rm Im} G$ in our theory could
only result from a negative residual. However, the graph of
$\Sigma$ at Matsubara frequencies for all $k$-points is
qualitatively similar to whose shown in the upper panel of Figure
\ref{Fig7perc}. It is obvious these graphs have a negative
derivative at Fermi energy, so that the residual $Z= (i-{\rm}
\frac{\partial \Sigma}{\partial \omega})^{-1}$ must be positive.

\section{Conclusions}

To summarize, the transformation to dual fermion variables
completely reconstructs perturbation theory, starting with the
zeroth-order approximation which is accurate in the limits of both
very weak and very strong interactions. As a result, taking into
account just a few lower-order diagrams gives quite satisfactory
results, without having to resumm infinite series of diagrams.
Starting with DMFT as the best local approximation, we are able to
take into account nonlocal corrections in a regular perturbative
way. In contrast with several cluster approaches the method is
exactly translationally invariant and allows us to analyze how
different parts of the reciprocal space are distinctly affected by
correlation effects.

This approach can be setup either in phases with long-range order
(antiferromagnetism, superconductivity) or in phases without
long-range order (normal state) by not allowing for symmetry
breaking. The present article mostly deals with the latter case.
By doing so, we could focus on physical effects that are not
directly related to incipient long-range order. In particular, we
showed that the anisotropic destruction of quasiparticles and the
Fermi surface (at least, as presented in Figs.
\ref{FigSigmaNAN}-\ref{FigCutDisp}) is not due to precursor
effects of antiferromagnetism (or superconductivity) as soon as
the intermediate and strong coupling regimes are entered. Indeed,
it is associated with quite-short range physics, as illustrated by
the fact that only the short-range components of the self-energy
are found to have significant magnitude. This observation also
provides some support to cluster extensions of DMFT.

Although the destruction of quasiparticles in a momentum-selective
way is adequately captured by this approach and associated with
short-range correlations, more work is required (possibly
including symmetry breaking and incipient long-range order) in
order to reach a proper description of the pseudogap formation and
of its dependence on the doping level and on the $t'/t$ ratio.

The work was supported by RFFI (grants 08-02-01020, 08-03-00930,
08-02-91953), DFG (grant 436-RUS-113-938-0), FOM (The
Netherlands), CNRS and Ecole Polytechnique.

\appendix
\section{Exact relations for high-order cumulants}

Similarly to the exact relationship (\ref{exact}) between the initial and dual Green's function,
the one-to-one correspondence between higher-order momenta for the initial 
and dual system can be established. Particularly, the formula for the four-order 
Green's function was presented and discussed previously \cite{dualsusc}.
It was shown that the two-particle excitations
in the original and dual system are identical.
Here, we use the generating functional approach, that allows us to establish 
the general structure of such relationships for high momenta, and extend the conclusion 
about the two-particle excitations to all collective excitations, involving an arbitrary number of particles.

We start from the expression for action (\ref{Scf}), which includes both initial and dual variables.
Then we introduce the independent variations of initial and dual energy:
\begin{equation}
S[c, c^*, f, f^*; u, v]=S[c, c^*, f, f^*]+u_{12} c^*_1 c_2+v_{12} f^*_1 f_2,
\end{equation}
where $u$ and $v$ are infinitesimal and a summation over repeating indices is implied.

One can see that Taylor series of the functional 
\begin{equation}
F[u, v] = \ln \int e^{-S[c, c^*, f, f^*; u, v]}{\cal D} f {\cal D} f^* {\cal D} c {\cal D} c^*
\end{equation}
with powers of $u$ and $v$ correspond, respectively, 
to the cumulants of initial and dual system.
We remind that the second-order cumulant is the Green's function,
and higher-order cumulants are proportional to corresponding vertex parts. For example,
the fourth-order cumulant is $\frac{\partial^2 F}{\partial u_{3'2'} \partial u_{4'1'}}=X_{1234}-G_{23} G_{14} + G_{13} G_{24}$ ($X$ is the two-particle Green's function), whereas the fourth-order vertex 
$\Gamma^{(4)}_{1234}=G_{11'}^{-1} G_{22'}^{-1} \frac{\partial^2 F}{\partial u_{3'2'} \partial u_{4'1'}} G_{3'3}^{-1} G_{4'4}^{-1}$.

To establish a relation between the cumulants, let us integrate over 
$f^*, f$ in the previous formula. We obtain
\begin{equation}
\begin{array}{l}
F[u, v] = F_0[u, v] + \ln \int e^{-S[c, c^*; u, v]} {\cal D} c {\cal D} c^*\\  \\
F_0[u, v]=-\ln \det || {\rm I} +  (\Delta-\epsilon) \alpha^{-1} v \alpha^{-1}||\\   \\
 S[c, c^*; u, v]=S[c, c^*]+\Delta_\omega c^*_{\omega k \sigma} c_{\omega k \sigma} +\left( u_{12} -M_{12}  \right)c^*_1 c_2\\  \\
M=\left((\Delta-\epsilon)^{-1}+\alpha^{-1} v \alpha^{-1}\right)^{-1} 
\end{array}
\end{equation} 
Symbol ${\rm I}$ in the second line is the matrix unity, and the second 
term is the product of the corresponding matrices. The fourth line reads similarly.

Last expressions clearly show that the derivatives of $F[u, v]$ with respect to
$u$ and $v$ are related. A comparison of the first derivatives, for example, allows to reproduce formula (\ref{exact}). The last term of (\ref{exact}) comes from the 
differentiation of $F_0$.

Let us consider the fourth-order cumulants  $\frac{\partial^2 F}{\partial u_{32} \partial u_{41}}$ and $\frac{\partial^2 F}{\partial v_{32} \partial v_{41}}$. First of all we note that neither indices $1$ and $2$, not $3$ and $4$ should coinside, because overvise 
both cumulants vanish due the Fermi-operator algebra. For the case of different indices,
the differentiation is quite simple and gives, after 
putting $\alpha=g^{-1}$, formula (29) of the paper \onlinecite{dualsusc}.
\begin{equation}
\frac{\partial^2 F}{\partial u_{32} \partial u_{41}}=
L_{11'} L_{22'} \frac{\partial^2 F}{\partial v_{3'2'} \partial v_{4'1'}} R_{3'3} R_{4'4}.
\end{equation}
Here $L$ and $R$ are matrix inverse of $(\Delta-\epsilon)^{-1} g$ and $g (\Delta-\epsilon)^{-1}$, respectively.
It should be  emphasized that this expression 
does not contain any extra additive terms, in contrast to formula (\ref{exact}).
Formally this is because the second derivative $\frac{\partial^2 F_0}{\partial v_{32} \partial v_{41}}$ vanish, as one can check straightforwardly.
Physically this means that the two-particle excitations
in the original and dual system are identical \cite{dualsusc}.

There might be also instructive to re-express the last formula in terms of vertex
function. Putting also $\alpha=g^{-1}$, one obtains
\begin{equation}\label{exactGamma}
     \Gamma_{1234}=L_{11'}' L_{22'}' \Gamma^{dual}_{1'2'3'4'} R_{3'3}' R_{4'4}',
\end{equation}
where $L'=(1+\Sigma_{dual} g)^{-1}$ and $R'=(1+g \Sigma_{dual})^{-1}$.

One can see that the obtained formulas are formally valid also for the case of coinsiding indices, when both left- and right-hand sides vanish.

An advantage of the presented approach is that the derivation of the formulas for 
six and higher-order vertex parts appears to be literally the same as for the fourth order.
All the argumentation about the absence of the coinciding indices and vanishing of the high
derivatives of $F_0$ is valid for that case. Therefore formula 
(\ref{exactGamma}) is valid for vertex parts of any order,just a number of indices and multipliers $L', R'$ should be changed. From the physical point of view, we conclude that all collective excitations of the initial and dual ensemble are the same.

\section{Functional minimization, relation to DMFT, and
self-consistency condition}

It is clear from the present consideration that a proper choice of
the hybridization function $\Delta$ is crucial. A
functional-minimization scheme is suitable to clarify this issue.
Let us introduce a trial action $\tilde{S}[f,f^*]$. For clarity,
we put the subscript $\tilde{S}$ at the triangle brackets in this
section, to emphasize the the averaging is over the system with
trial action $\tilde{S}$. We consider Feynman's variational
functional
\begin{equation}\label{Min}
\begin{array}{c}
    <\tilde{S}>_{\tilde{S}}+\ln \int e^{-\tilde{S}} {\cal D} f{\cal
    D}f^*- \\  \\
    -<S>_{\tilde{S}}-\ln \int e^{-S} {\cal D} f{\cal D} f^*=max.
\end{array}
\end{equation}
A straightforward variation $\tilde{S}\to\tilde{S}+\delta S$ gives
an extremum condition
\begin{equation}\label{extremum}
    <(S-\tilde{S})\delta S>_{\tilde{S}}=<(S-\tilde{S})>_{\tilde{S}}<\delta
    S>_{\tilde{S}}.
\end{equation}
For an arbitrary $\delta S$, this indeed means that the extremum
of (\ref{Min}) is delivered by $\tilde{S}=S$, up to an additive
constant. In this case (\ref{Min}) vanishes. The larger value of
(\ref{Min}) corresponds to the better approximation.

There is an important point:  since dual action depends on
$\Delta$, the condition (\ref{Min}) can be used to determine the
optimal $\Delta$. The variation with respect to $\Delta$ gives
\begin{equation}\label{Deltaexact1}
    \frac{\delta <S>_{\tilde{S}}}{\delta \Delta}=0.
\end{equation}
Here we took into account that variations of $\tilde{S}$ and
$\Delta$ are independent, so the first two terms of (\ref{Min}) do
not vary with $\Delta$. As for the last term, it is exactly $\ln
Z$ and therefore independent of $\Delta$ as well.

 Now, recalling $S[f,f^*]=-\ln \int
e^{-S[c,c^*,f,f^*]}$ and substituting (\ref{Scf}), we obtain after
certain transformations that (\ref{Deltaexact1}) corresponds to
the condition
\begin{equation}\label{Deltaexact2}
\begin{array}{c}
    G_{\omega, r=0}=<g^{imp}[f_i, f^*_i]>_{\tilde{S}},\\
    \\
    g^{imp}[f_i, f^*_i]=\frac{\int c^*_{\omega i} c_{\omega i}
    e^{-S_{site}[c_i c^*_i f_i f^*_i]} {\cal D} c^*_i {\cal D} c_i}
    {{\int e^{-S_{site}[c_i c^*_i f_i f^*_i]} {\cal D} c^*_i {\cal D}
    c_i}}.
\end{array}
\end{equation}
Here $S_{site}$ is defined by formula (\ref{Ssite}) and
$G_{r=0}=N^{-1} \sum_k G_{k}$ is local part of the Green's
function. While deriving these formulas, it is useful to take into
account that $\alpha=g^{-1}$ is just a scaling factor standing at
$f^*, f$, and there is no need to vary this quantity: one can vary
with respect to $\Delta$ at fixed $\alpha$ and put $\alpha=g^{-1}$
afterwards.

 Actually, the criterion (\ref{Deltaexact2}) has a very clear meaning: local
part of the Green's function equals the Green's function of the
single-site action $S_{site}$, averaged over the fluctuations of
$f$. Neglecting these fluctuations, one obtains just a DMFT
condition for hybridization function, that is $G_{\omega,
r=0}=g_\omega$.

To make the consideration more clear, let us first consider the
Gaussian approximation for dual variables, $\tilde{S}=-{\cal
G}_{dual}^{-1} f^* f$. Let us show  for this Gaussian trial
action, the DMFT condition
\begin{equation}\label{DMFTcond}
    {\cal G}_{\omega,r=0}=g_\omega
\end{equation}
satisfies (\ref{Deltaexact2}) {\it exactly} (call this statement
T1). The proof is based on the observation that the condition
(\ref{DMFTcond}) is equivalent to the requirement that the local
part of dual Green's function equals zero,
\begin{equation}\label{DMFTconddual}
   {\cal G}_{r=0}^{dual}=0,
\end{equation}
as one can easily check with formulas (\ref{DMFTdual},
\ref{DMFT}). Further, since $\tilde{S}$ is Gaussian, formula
(\ref{DMFTconddual}) means that all local momenta $<f_i^* f_i>$,
$<f_i^* f_i f_i^* f_i>, ...$ equal zero. It means that local
fluctuations of $f, f^*$ are virtually absent, therefore
$<g^{imp}_\omega[f,f^*]>=g_\omega$ and (\ref{Deltaexact2}) becomes
(\ref{DMFTcond}). To obtain a formal proof, one should consider an
average of the Taylor series for $g^{imp}[f,f^*]$. These series
starts from $g_\omega$, whereas the average of any higher term
vanishes. This profs T1.

Next, it is possible also to show that the DMFT Green's function
is optimal with respect to the variations of the Gaussian trial
action (call this statement T2). With a variation
$\tilde{S}=-{\cal G}_{dual}^{-1} f^* f\to\tilde{S}=-{\cal
G}_{dual}^{-1} f^* f+0 f_1^* f_2$, formula (\ref{extremum})
becomes
\begin{equation}\label{DMFTextremum}
    <(S+{\cal G}_{dual}^{-1} f^* f) f_1^* f_2>_{\tilde{S}}=
    <S + {\cal G}_{dual}^{-1} f^* f>_{\tilde{S}} <f_1^* f_2>_{\tilde{S}}.
\end{equation}
The essential point is again that since all local momenta of $f,
f^*$ are vanished because of (\ref{DMFTconddual}), and the dual
potential $V$ is local in space, all the nonlinearity  drops out
from the (\ref{DMFTextremum}). It means that both left- and right-
hand sides of (\ref{DMFTextremum}) equal the same value, if
$-{\cal G}_{dual}^{-1} f^* f$ equals the Gaussian part of the dual
action. This proofs T2.

So, we have shown that the DMFT procedure can be considered as the
Gaussian approximation for the dual variables, which is optimal in
sense of Feynman minimization criterion, with respect to both
trial action and hybridization function.

Beyond the Gaussian trial action, an analytical treatment of the
extremal criterion (\ref{Min}) is hardly possible. Therefore, in
the main body of the theory we treat the dual system
perturbatively, using the diagram series with respect to $V$ and
the hybridization function defined from the condition
(\ref{NoLoops}).

\end{document}